\documentstyle[aps,12pt]{revtex}
\newcommand{\bb}{\begin{equation}}
\newcommand{\ee}{\end{equation}}
\newcommand{\ba}{\begin{array}}
\newcommand{\ea}{\end{array}}

\begin {document}

\title{Scattering by a contact potential in three
       and lower dimensions\thanks
       {published in J. Phys. A {\bf 34} (2001) 1911-1918.}}
\author{Qiong-gui Lin\thanks{E-mail addresses: qg\_lin@163.net,
        qg\_lin@263.net}}
\address{China Center of Advanced Science and Technology (World
    Laboratory),\\
        P.O.Box 8730, Beijing 100080, People's Republic of China
        \thanks{not for correspondence}\\
        and\\
        Department of Physics, Zhongshan University, Guangzhou
        510275,\\
        People's  Republic of China}

\maketitle
\vfill

\begin{abstract}
{\normalsize We consider the scattering of nonrelativistic particles
in three dimensions by a contact potential $\Omega\hbar^2\delta(r)/
2\mu r^\alpha$ which is defined as the $a\to 0$ limit of
$\Omega\hbar^2\delta(r-a)/2\mu r^\alpha$. It is surprising that it
gives a nonvanishing cross section when $\alpha=1$ and $\Omega=-1$.
When the contact potential is approached by a spherical square well
potential instead of the above spherical shell one, one obtains
basically the same result except that the parameter $\Omega$ that
gives a nonvanishing cross section is different. Similar problems in
two and one dimensions are studied and results of the same nature are
obtained.}
\end{abstract}
\vfill

\leftline {PACS number(s): 03.65.Nk}
\newpage
\baselineskip 15pt

\section{Introduction}                   
In this paper we consider a very simple problem, the scattering of
nonrelativistic particles by a contact potential, which is nonzero
(in general infinite) at one point and vanishes elsewhere. More
specifically, in three dimensions we consider the potential \bb
V(r)={\Omega\hbar^2 \over 2\mu}{\delta(r)\over r^\alpha},
\ee     
where $\Omega$ and $\alpha$ are real parameters and $\mu$ is the
mass of the particle.
Since this is somewhat difficult to handle in practice, we are
actually considering the $a\to 0$ limit of the spherical shell
potential
\bb
V_a(r)={\Omega\hbar^2\over 2\mu}{\delta(r-a)\over r^\alpha}
={\Omega\hbar^2\over 2\mu}{\delta(r-a)\over a^\alpha}
\ee     
where $a>0$, which is better defined. The latter is a contact
potential since in the limit it is nonvanishing (and infinite if
$\alpha\ge 0$) only at $r=0$, though one may think it is different
from the former. The interest of such a problem is threefold. First,
according to classical mechanics, the scattering cross section for a
contact potential in three or two dimensions is obviously zero. In
one dimension an incident particle would be bounced back by a contact
potential (total reflection) if the potential is infinite at the
nonvanishing point. However, the case may be different in quantum
mechanics. Second, if a finite cross section is possible in quantum
mechanics, one may think that a ``larger'' infinity (higher
singularity) at the nonvanishing point would give a larger cross
section, but this turns out to be incorrect. Only a ``proper''
infinity gives a nonvanishing result. For the above potential, a
nonvanishing cross section is obtained only when $\alpha=1$ and
$\Omega=-1$. Third, an exact result is easily available. This is
because all phase shifts except $\delta_0$ vanish in the limit $a\to
0$, which is what is expected for a contact potential.

In the next section we consider the above problem in three
dimensions. Instead of the spherical shell potential (2), the contact
potential (1) can be approached by some other potential as well. It
is then of interest to see whether the same result is obtained. A
simple candidate is the square well potential. This is also
considered. It turns out that the result is basically the same,
except that the parameter $\Omega$ that gives a nonvanishing cross
section is different.

In section III we consider a similar problem in two dimensions. It
turns out that the potential of the same form as in three dimensions
gives a vanishing cross section for any $\alpha$ and $\Omega$. We
then consider the $a\to 0$ limit of the circular ring potential
\bb
V_a(r)={\Omega\hbar^2 \over 2\mu}{\delta(r-a)\over r^\alpha
[-\ln(r/a_0)]^{\beta}} ={\Omega\hbar^2 \over 2\mu}{\delta(r-a)\over
a^\alpha[-\ln(a/a_0)]^{\beta}},
\ee     
where $a_0$ is a length scale and $\beta$ is another real parameter.
We take $a_0>a$ (note that we would finally take the limit $a\to 0$)
such that $-\ln(a/a_0)$ is positive. It turns out that a nonvanishing
cross section is obtained only when $\alpha=1$, $\beta=1$ and
$\Omega=-1$. We also consider a square well potential that approaches
the same contact potential. It turns out that all results are exactly
the same as for the above circular ring potential. This is somewhat
different from the case in three and one dimensions. In section IV we
consider the $a\to 0$ limit of the double delta potential in one
dimension:
\bb
V_a(x)={\Omega\hbar^2\over
2\mu}{\delta(x-a)+\delta(x+a)\over |x|^\alpha} ={\Omega\hbar^2\over
2\mu}{\delta(x-a)+\delta(x+a)\over a^\alpha},
\ee     
where $a>0$. When $\alpha=0$, this tends to the ordinary $\delta(x)$
potential and we indeed get the known result [1], which is already
different from the classical one. Here we are interested in the
results for other values of $\alpha$. It turns out that $\alpha<0$
leads to total transmission. This is an expected result since there
is no singularity in this case. When $\alpha>0$ one has in general
total reflection. This is also expected since the singularity is
strong. What is unexpected is, however, when $\alpha=1$ and
$\Omega=-1$ one has total transmission with a phase shift $\pi$. When
the contact potential is approached by a square well potential, the
result is roughly the same. However, in addition to the total
transmission with a phase shift $\pi$ for some specific values of
$\Omega$ (not $-1$), we have true total transmission for some other
values of $\Omega$.

\section{Three dimensions}     
Consider scattering in three dimensions by
the spherically symmetric potential (2). If $a$ is a constant, this
is the ordinary spherical shell potential studied in the literature
[1], and the parameter $\alpha$ plays no role. However, here we are
interested in the limit $a\to 0$, so different $\alpha$ would lead to
different results. The radial equation for the $l$th partial wave is
\bb R''(r)+{2\over r}R'(r)+\left[k^2-{l(l+1) \over r^2}\right]R(r)
-{\Omega\over a^\alpha}R(a)\delta(r-a)=0, \quad l=0,1,2,\ldots,
\ee     
where $k=\sqrt{2\mu E}/\hbar$ and $E>0$ is the energy of the incident
particle. The solution of this equation is
$$
R(r)=A_l j_l(kr), \quad r<a, \eqno(6{\rm a})$$
$$
R(r)=\cos\delta_l j_l(kr)-\sin\delta_l n_l(kr), \quad r>a,
\eqno(6{\rm b})$$ where $\delta_l$ is the phase shift, $A_l$ is a
constant, and $j_l$ and $n_l$ are the spherical Bessel and Neumann
functions, respectively. The two parts of the solution must be
appropriately connected at $r=a$. The connection conditions are
$$
R(a^+)=R(a^-), \eqno(7{\rm a})$$
$$
R'(a^+)-R'(a^-)={\Omega\over a^\alpha}R(a),
\eqno(7{\rm b})$$
where $a^\pm=a\pm 0$. Eq. (7b) is obtained by integrating Eq. (5)
from $a^-$ to $a^+$. Since the jump of $R'(r)$ at $r=a$ is finite
we have Eq. (7a). The phase shift is determined by these conditions
and the result is
\addtocounter{equation}{2}
\bb
\tan\delta_l={b\xi j_l^2(\xi)\over
-\xi^{\alpha-1}+b\xi j_l(\xi)n_l(\xi)},
\ee     
where $\xi=ka$ and $b=\Omega k^{\alpha-1}$. The scattering amplitude
is given in many textbooks of quantum mechanics as
\bb
f(\theta)={1\over 2ik}\sum_{l=0}^\infty (2l+1)
[\exp(2i\delta_l)-1]P_l(\cos\theta).
\ee     
Though we have an analytic expression for the phase shifts, it is
difficult to have a closed result for the scattering amplitude since
the summation in Eq. (9) cannot be worked out. Here we are only
interested in the limit $a\to 0$ (or $\xi\to 0$), however, so the
problem is much simpler. If $\alpha>1$, the first term in the
denominator of Eq. (8) can be neglected in comparison with the
second. So we have obviously $\tan\delta_l\to 0$ for all $l$ and
$f(\theta)\to 0$. If $\alpha<1$ (including negative $\alpha$), the
second term in the denominator can be neglected in comparison with
the first, and the same result is easily achieved. If $\alpha=1$, the
two terms in the denominator are of the same order. One should be
careful in this case. Obviously, the numerator behaves like
$\xi^{2l+1}$, while the denominator behaves like $\xi^0$ or, if $b$
takes some specific value, like $\xi^2$. So If $l\ge 1$, we have
$\tan\delta_l\to 0$. A nonvanishing phase shift is therefore possible
only when $l=0$, as expected for a contact potential. If $l=0$ and
$b\ne -1$, we have $\tan\delta_0\to 0$ as well, and thus
$f(\theta)\to 0$. If $b=-1$, however, we have $\tan\delta_0\sim
3/2\xi\to \infty$, so that $\delta_0=\pi/2,~{\rm mod}~\pi$. Only in
this case do we have a nonvanishing scattering amplitude \bb
f(\theta)={i\over k},
\ee     
which is of course independent of $\theta$. The differential and
total cross sections are
\bb
\sigma(\theta)={1\over k^2}={\hbar^2\over \mu^2 v^2},\quad
\sigma_t={4\pi \over k^2}={4\pi\hbar^2 \over \mu^2 v^2},
\ee     
where $v$ is the classical velocity of the incident particle.
Therefore only a contact potential with proper singularity and
proper strength gives a nontrivial result. We make some remarks.
First, the result is inversely proportional to the incident energy.
Second, $b=-1$ yields $\Omega=-1$ since $\alpha=1$. Third,
when $\alpha=1$, $\tan\delta_0$ depends only on $\xi=ka$ (not
on $k$ and $a$ separately), so that
$\lim_{a\to 0}\tan\delta_0=\infty$ for finite $k$ implies
$\lim_{k\to 0}\tan\delta_0=\infty$ for finite $a$.
According to the analysis of the Levinson theorem [2-5],
one has in general $\lim_{k\to 0}\tan\delta_0=0$,
only when there exists a half-bound state [6] with zero energy
and $l=0$ does the limit become infinite.
In fact, the zero-energy solution with $l=0$ to Eq. (5) is
$$
R(r)=1, \quad r<a, \eqno(12{\rm a})$$
$$
R(r)=B_0 r^{-1}, \quad r>a, \eqno(12{\rm b})$$
where $B_0$ is a
constant. The two parts of the solution can be connected by the
conditions (7) only when $\Omega=-1$. This is consistent with the
above result. The solution (12) is called a half-bound state because
it tends to zero at infinity but is not normalizable.

As pointed out in section I, the contact potential can be approached
by some potential other than the above spherical shell one. It is
then of interest to see whether the same conclusion is achieved. We
consider the spherical square well potential ($\Omega<0$ is assumed
for such potentials in all dimensions)
\addtocounter{equation}{1}
\begin{equation}
V^a(r)=\left\{
\begin{array}{ll}
3\Omega\hbar^2/2\mu a^{\alpha+1},\quad & r<a, \\
0, & r>a.
\end{array}\right.
\end{equation}  
This gives the integration $\int V^a(r)\;d^3 x=2\pi\Omega\hbar^2/\mu
a^{\alpha-2}$, the same as given by the spherical shell potential
(2). Thus when $a\to 0$ they would tend to the same contact
potential. The phase shifts for this potential can be easily found.
They are determined by
\begin{equation}
 \tan\delta_l={\eta j_l(\xi)j_{l+1}(\eta)
 -\xi j_{l+1}(\xi)j_l(\eta)\over
 \eta n_l(\xi)j_{l+1}(\eta)
 -\xi n_{l+1}(\xi)j_l(\eta)},
\end{equation}  
where $\xi$ is defined as before, and
$\eta=\sqrt{\xi^2-3b\xi^{1-\alpha}}$. Then we let $\xi\to 0$ and
analyze the behavior of the phase shifts. After careful calculations,
it is found that $\tan\delta_l\to 0$ for all $l$ if $\alpha\ne 1$.
Thus a nontrivial result is possible only when $\alpha=1$, the same
conclusion as before. If $\alpha=1$, it can be shown that
$\tan\delta_l\to 0$ for all $l\ne 0$. As for $l=0$, we have in
general $\tan\delta_0\to 0$, except when $j_{-1}(\sqrt{-3b})=0$ or
$n_{0}(\sqrt{-3b})=0$. In the latter case we have
$\tan\delta_0\sim\xi^{-1}\to\infty$, and hence the nontrivial results
(10) and (11). The above condition is equivalent to
$\cos(\sqrt{-3b})=0$ or (note that $\alpha=1$)
\begin{equation}
\Omega=-{(2N-1)^2\pi^2\over 12},
\end{equation}  
where $N$ is a natural number. The difference from the previous case
is that for the spherical shell potential there is only one specific
value of $\Omega$ that leads to a nontrivial cross section while in
the present case we obtain a series of such values (but the condition
$\alpha=1$ is the same). This difference is related to the different
situation for the appearance of half-bound states with $l=0$ (for
$\alpha=1$ and finite $a$). For the spherical shell potential the
half-bound state can appear only once when $\Omega=-1$, while for the
square well potential it can appear many times when $\Omega$ takes on
the above values.

\section{Two dimensions}       
In this section we consider a similar problem in two dimensions. We
still use ($r$, $\theta$) for the polar coordinates on the $xy$
plane. These should not be confused with those in three dimensions.
If one considers a potential of the form (2), it can be shown that
the result is trivial in the limit $a\to 0$. So we consider the
potential (3). The radial wave equation for the $m$th partial wave is
\bb
R''(r)+{1\over r}R'(r)+\left(k^2-{m^2 \over r^2}\right)R(r)
-{\Omega\over a^\alpha[-\ln(a/a_0)]^\beta}R(a)\delta(r-a)=0, \quad
m=0,1,2,\ldots.
\ee     
The solution of this equation is
$$
R(r)=C_m J_m(kr), \quad r<a, \eqno(17{\rm a})$$
$$
R(r)=\cos\delta_m J_m(kr)-\sin\delta_m N_m(kr), \quad r>a,
\eqno(17{\rm b})$$
where $\delta_m$ is the phase shift, $C_m$ is a
constant, and $J_m$ and $N_m$ are the ordinary Bessel and Neumann
functions, respectively. The connection conditions at $r=a$ are
$$
R(a^+)=R(a^-), \eqno(18{\rm a})$$
$$
R'(a^+)-R'(a^-)={\Omega\over a^\alpha [-\ln(a/a_0)]^\beta}R(a).
\eqno(18{\rm b})$$ The phase shift is determined by these conditions
and the result is \addtocounter{equation}{2} \bb \tan\delta_m={b\pi
J_m^2(\xi)\over -2\xi^{\alpha-1}[-\ln(\xi/\xi_0)]^\beta+b\pi
J_m(\xi)N_m(\xi)},
\ee     
where $\xi=ka$, $\xi_0=ka_0$ and $b=\Omega k^{\alpha-1}$.
The scattering amplitude is given by [7]
\bb
f(\theta)=-{i\over \sqrt{2\pi k}}\sum_{m=-\infty}^{+\infty}
[\exp(2i\delta_{|m|})-1]e^{im\theta}.
\ee     
As before we do not attempt to study the general case in detail, but
are interested in the limit $a\to 0$ (or $\xi\to 0$) only. If
$\alpha>1$, the first term in the denominator of Eq. (19) can be
neglected in comparison with the second, so we have obviously
$\tan\delta_m\to 0$ for all $m$ and $f(\theta)\to 0$. If $\alpha<1$
(including negative $\alpha$), the second term in the denominator can
be neglected in comparison with the first, and the same result is
easily achieved. Therefore a nontrivial result is possible only when
$\alpha=1$. In this case
\bb
\tan\delta_m={b\pi J_m^2(\xi)\over
-2[-\ln(\xi/\xi_0)]^\beta+b\pi J_m(\xi)N_m(\xi)}.
\ee     
If $m\ge 1$, the numerator behaves like $\xi^{2m}$, while the
denominator behaves like $\xi^0$ if $\beta<0$ or like
$[-\ln(\xi/\xi_0)]^\beta$ if $\beta>0$ (note that the case $\beta=0$
leads to a trivial result which has been mentioned at the beginning
of this section), so if $m\ge 1$, we have $\tan\delta_m\to 0$. A
nonvanishing phase shift is therefore possible only when $m=0$, as
expected for a contact potential. If $m=0$, the numerator behaves
like $\xi^{0}$, while the denominator behaves like $\ln\xi$ if
$\beta<1$ or like $[-\ln(\xi/\xi_0)]^\beta$ if $\beta>1$.  So if
$\beta\ne 1$ we have $\tan\delta_0\to 0$ as well, and thus
$f(\theta)\to 0$. If $\beta=1$ but $b\ne -1$, the denominator still
behaves like $\ln\xi$, so we still have $\tan\delta_0\to 0$, and thus
$f(\theta)\to 0$. However, if $\beta=1$ and $b=-1$, we have in the
limit $a\to 0$ \bb \tan\delta_0={\pi\over 2\gamma+2\ln(\xi_0/2)},
\ee     
where $\gamma$ is Euler's constant, and in this case
we have a nonvanishing scattering amplitude
\bb
f(\theta)=\sqrt{2\pi \over k}{1\over 2\gamma+2\ln(\xi_0/2)-i\pi},
\ee     
which is of course independent of $\theta$, but note that the result
depends on $a_0$ or $\xi_0$. The differential and total cross
sections are
\bb
\sigma(\theta)={2\pi \over
k[4(\gamma+\ln(\xi_0/2))^2+\pi^2]} ={2\pi\hbar \over \mu
v[4(\gamma+\ln(\xi_0/2))^2+\pi^2]},
\ee     
\bb
\sigma_t={4\pi^2 \over k[4(\gamma+\ln(\xi_0/2))^2+\pi^2]}
={4\pi^2 \hbar\over \mu v[4(\gamma+\ln(\xi_0/2))^2+\pi^2]}.
\ee     
Therefore only when $\alpha=1$, $\beta=1$, and $\Omega=-1$ do we get
a nontrivial result.

As in three dimensions, we then consider the two-dimensional square
well potential
\begin{equation}
V^a(r)=\left\{
\begin{array}{ll}
\Omega\hbar^2/\mu a^{\alpha+1}[-\ln(a/a_0)]^\beta,\quad & r<a, \\
0, & r>a.
\end{array}\right.
\end{equation}  
This gives the integration $\int V^a(r)\;d^2 x=\pi\Omega\hbar^2/\mu
a^{\alpha-1}[-\ln(a/a_0)]^\beta$, the same as given by the circular
ring potential (3). Thus when $a\to 0$ they would tend to the same
contact potential. The phase shifts for this potential can be easily
found. They are determined by
\begin{equation}
 \tan\delta_m={\eta J_m(\xi)J_{m+1}(\eta)
 -\xi J_{m+1}(\xi)J_m(\eta)\over
 \eta N_m(\xi)J_{m+1}(\eta)
 -\xi N_{m+1}(\xi)J_m(\eta)},
\end{equation}  
where $\xi$ is defined as before, and
$\eta=\sqrt{\xi^2-2b\xi^{1-\alpha}[-\ln(\xi/\xi_0)]^{-\beta}}$. Then
we let $\xi\to 0$ and analyze the behavior of the phase shifts. The
present situation is somewhat more complicated than that in three
dimensions. The various cases with different parameter values should
be discussed separately. After careful calculations, it is found that
the nontrivial results (22-25) are obtained only when $\alpha=1$,
$\beta=1$, and $\Omega=-1$. This is exactly the same conclusion as
for the circular ring potential. Note that the nontrivial result in
two dimensions has nothing to do with the existence of half-bound
states. Thus the situation is rather different from that in three or
one dimensions (see below).

\section{One dimension}         
Now we turn to the potential (4) in one dimension. For scattering of
particles incident from the left, the boundary conditions are
$$
\psi(x)=e^{ikx}+R e^{-ikx}, \quad  x<-a, \eqno(28{\rm a})$$
$$
\psi(x)=T e^{ikx}, \quad  x>a. \eqno(28{\rm b})$$ Here $R$ and $T$
are the reflection and transmission amplitudes, respectively. On the
other hand, with a given energy (or a given $k$), there exist two
linearly independent solutions to the Schr\"odinger equation. Since
the potential is an even function of $x$, one can choose the two
solutions to have definite parity. The even-parity solution has the
form
$$
\psi_{+}(x)=A_{+}\cos kx, \quad  |x|<a, \eqno(29{\rm a})$$
$$
\psi_{+}(x)=\cos(k|x|+\delta_{+}), \quad  |x|>a, \eqno(29{\rm b})$$
while the odd-parity one has
$$
\psi_{-}(x)=A_{-}\sin kx, \quad  |x|<a, \eqno(30{\rm a})$$
$$
\psi_{-}(x)=\epsilon(x)\sin(k|x|+\delta_{-}), \quad  |x|>a,
\eqno(30{\rm b})$$
where $A_\pm$ are constants, $\epsilon(x)$ is
the sign function, and $\delta_\pm$ are phase shifts. A scattering
solution can be written as a linear combination of $\psi_\pm$, so we
have
$$
R=\textstyle\frac 12(e^{i2\delta_+}-e^{i2\delta_-}),\eqno(31{\rm
a})$$
$$
T=\textstyle\frac 12(e^{i2\delta_+}+e^{i2\delta_-}).\eqno(31{\rm
b})$$
The phase shifts are determined by the connection condition at
$x=a$:
$$
\psi(a^+)=\psi(a^-), \eqno(32{\rm a})$$
$$
\psi'(a^+)-\psi'(a^-)={\Omega\over a^\alpha}\psi(a). \eqno(32{\rm
b})$$
The result is
$$
\tan\delta_+={b\cos^2\xi\over -\xi^\alpha+b\sin\xi\cos\xi},
\eqno(33{\rm a})$$
$$
\tan\delta_-=-{b\sin^2\xi\over \xi^\alpha+b\sin\xi\cos\xi},
\eqno(33{\rm b})$$
where $\xi=ka$ and $b=\Omega k^{\alpha-1}$. Then
we consider the limit $a\to 0$ (or $\xi\to 0$). If $\alpha=0$, we
have $\tan\delta_+=-b=-\Omega/k$ and $\tan\delta_-=0$. This leads to
the known result for an ordinary $\delta(x)$ potential [1]. If
$\alpha<0$, we have $\tan\delta_+=\tan\delta_-=0$, so that $R=0$,
$T=1$, that is, total transmission. This is an expected result since
there is actually no singularity in this case.  If $\alpha>0$, we
have in general $\tan\delta_+=\infty$ and $\tan\delta_-=0$, so that
$R=-1$, $T=0$, that is, total reflection with a phase shift $\pi$
(this is not refered to the above $\delta_\pm$). This is also an
expected result since the singularity is strong in this case.
However, if $\alpha=1$ and $b=-1$ (or $\Omega=-1$), we have
$\tan\delta_+=\infty$ and $\tan\delta_-=\infty$, so that $R=0$,
$T=-1$, that is, total transmission with a phase shift $\pi$. This is
an unexpected result since it is among the cases with strong
singularity. We point out that there exists a half-bound state with
zero energy and odd parity in this case (for any finite $a$).

As in three and two dimensions, we now consider the square well
potential
\addtocounter{equation}{6}
\begin{equation}
V^a(x)=\left\{
\begin{array}{ll}
\Omega\hbar^2/2\mu a^{\alpha+1},\quad & |x|<a, \\
0, & |x|>a.
\end{array}\right.
\end{equation}  
This gives the integration $\int V^a(x)\;d x=\Omega\hbar^2/\mu
a^{\alpha}$, the same as given by the double delta potential (4).
Thus when $a\to 0$ they are expected to tend to the same contact
potential. The phase shifts for this potential are determined by
$$
\tan\delta_+={\eta\tan\eta-\xi\tan\xi\over \xi+\eta\tan\xi\tan\eta},
\eqno(35{\rm a})$$
$$
\tan\delta_-={\xi\tan\eta-\eta\tan\xi\over \eta+\xi\tan\xi\tan\eta},
\eqno(35{\rm b})$$ where $\xi$ is defined as before, and
$\eta=\sqrt{\xi^2-b\xi^{1-\alpha}}$. Then we let $\xi\to 0$ and
analyze the behavior of the phase shifts. After careful calculations,
it is found that they lead to expected results when $\alpha\ne 1$,
and when $\alpha=1$ but $\Omega$ does not take on specific values. An
unexpected result is obtained when $\alpha=1$ and $\Omega$ takes on
values from one of the two sets given below. The first nontrivial
result is $R=0$, $T=-1$, i.e., total transmission with a phase shift
$\pi$. This is the same as for the double delta potential, but it
happens when
\addtocounter{equation}{1}
\begin{equation}
\Omega=-{(2N-1)^2\pi^2\over 4},
\end{equation}  
where $N$ is a natural number, which is different from the previous
value. Note that there exists a half-bound state with odd parity (for
$\alpha=1$ and finite $a$) when $\Omega$ takes on any one of the
above values. In contrast, for the double delta potential such a
state can appear only once when $\Omega=-1$. The second nontrivial
result is $R=0$, $T=1$, i.e., true total transmission. This happens
when
\begin{equation}
\Omega=-N^2\pi^2,
\end{equation}  
where $N$ is a natural number. There is no similar result for the
double delta potential. Note that there exists a half-bound state
with even parity for the square well potential when $\Omega$ takes on
values from this sequence, while for the double delta potential there
exists no such state. Thus we see once again that the difference in
the unexpected result is related to the different situation for the
appearance of half-bound states.

\section*{Acknowledgments}

The author is grateful to Professor Guang-jiong Ni for encouragement.
This work was supported by the
National Natural Science Foundation of China.



\begin{thebibliography}{99}

\bibitem{1}S. Fl\"ugge, {\it Practical Quantum Mechanics}
          (Springer-Verlag, New York, 1974).

\bibitem{2}N. Levinson, K. Dan. Vidensk. Selsk. Mat.-Fys. Medd.
           {\bf 25}, No.9 (1949).

\bibitem{3}G.-J. Ni and S.-Q. Chen, {\it Levinson Theorem, Anomaly,
           and the Phase Transition of Vacuum} (Shanghai Scientific
           \& Technical, Shanghai, 1995) (in Chinese).

\bibitem{4}G.-J. Ni, Phys. Energ. Fort. Phys. Nucl. {\bf 3},
432 (1979).

\bibitem{5}Q.-G. Lin, Eur. Phys. J. D {\bf 7}, 515 (1999).

\bibitem{6}R. G. Newton, J. Math. Phys. {\bf 1}, 319 (1960).

\bibitem{7}Q.-G. Lin, Phys. Rev. A {\bf 56}, 1938 (1997).

\end{thebibliography}
\end{document}